\documentclass[10pt]{iopart}
\usepackage[dvipdfmx]{graphicx}
\usepackage{ulem}
\usepackage{here}
\def\vector#1{\mbox{\boldmath $#1$}}

\begin{document}
\title[Exactly-solvable model of solid-fluid phase transition 
for crystals with two particles in a primitive cell]{Generalization of exactly-solvable model to exhibit solid-fluid phase transition 
in crystal structures with two particles in a primitive cell}
\author{Hisato Komatsu}
\address{ Graduate school of Arts and Sciences, The University of Tokyo, 3-8-1 Komaba, Meguro-ku, Tokyo 153-8902, Japan \footnote{Present address: International Center
for Materials Nanoarchitectonics,  National Institute for Materials Science,  Tsukuba, Ibaraki 305-0044, Japan} }
\ead{KOMATSU.Hisato@nims.go.jp }

\begin{abstract}
In our previous paper [H.\ K., J.~Stat.~Mech.\ (2015) P08020], we investigated an interacting-particle model with infinite-range cosine potentials, and derived  the partition function which shows solid-fluid phase transition by exact calculation. However, we could  treat only simple lattice structures in which more than one stable point exist in a primitive cell such as the triangular or face-centered cubic lattice. In the present paper, we generalize our previous scheme to more complicated lattice structures with two particles in a primitive cell. Generalization to more complicated lattice structures is straightforward.
\end{abstract}

\section{Introduction \label{intro}}
Theoretical understanding of solid-fluid phase transitions from microscopic Hamiltonians is one of the fundamental problems of the condensed matter physics and statistical physics, and numerous theoretical investigations have been attempted so far~\cite{LJD,MOI,KirMon,RY,CA,DA,CL}. However, derivation of the thermodynamic quantities which shows phase transition by exact calculation has been quite limited~\cite{FP,CF,KH}. For example, Carmesin and Fan~\cite{CF} studied a one-dimensional model of interacting particles with a cosine and hard-core potentials, but their scheme to treat the hard-core potential was difficult to generalize to higher dimensions. 

In our previous paper~\cite{KH}, we treated $d$-dimensional classical particles interacting with each other with cosine potentials, and proved the existence of the solid-fluid phase transition by exact calculation of the partition function similarly to that of~\cite{CF}. We studied simple crystal structures containing one position for each particle in a primitive cell, such as the triangular, face-centered cubic (fcc), or body-centered cubic (bcc) lattices. However, such a scheme could not be applied to the models with more complicated crystal structures which has more than one position in a primitive cell. Previous theoretical approaches to the solid-fluid phase transition by investigating order parameters, such as the density functional theory, also faced this problem, and some trials overcame it. For example, Yussouff included the above effect by taking large reciprocal lattice vectors into account~\cite{Yussouff}. 

\begin{figure}[!hbp]
\begin{center}
\includegraphics[width = 8.0cm]{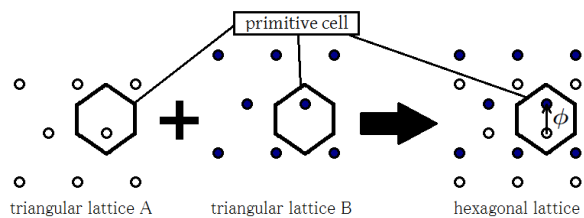}
\caption{Construction of hexagonal lattice from two triangular lattices: Hexagonal lattice is composed of two triangular lattices, and the displacement between these two sublattices is expressed as the constant vector $\vector{\phi}$. }
\label{hl_illust}
\end{center}
\end{figure}

\begin{figure}[!hbp]
\begin{center}
\includegraphics[width = 9.0cm]{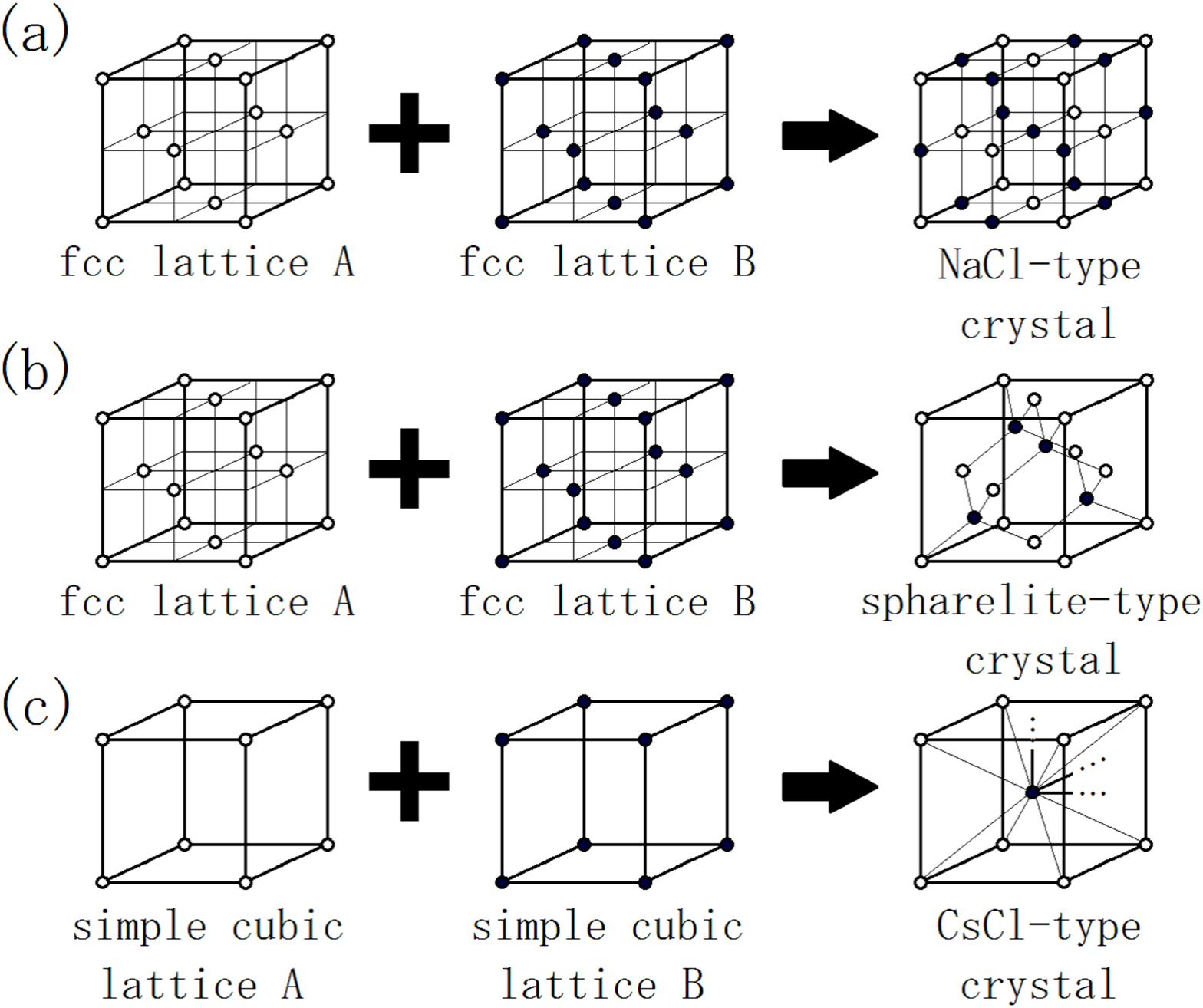}
\caption{Crystal structures composed of two sublattices: NaCl-type and sphalerite-type crystals are composed of two fcc lattices, and CsCl-type crystal of two three-dimensional simple cubic lattices. Difference between NaCl-type and sphalerite-type crystal is the displacement between these two sublattices. }
\label{crys_illust}
\end{center}
\end{figure}

In the present paper, we introduce the models of binary mixture whose solid phase includes two positions for particles in a primitive cell. Specifically, we consider a model containing two types of particles and each one forms a sublattice of the crystal. 
For example, the hexagonal lattice is composed of two sublattices of triangular lattices (figure \ref{hl_illust}),  the NaCl-type and sphalerite-type crystals of two fcc lattices (figures \ref{crys_illust}(a) and \ref{crys_illust}(b)), and the CsCl-type crystal of two three-dimensional simple cubic lattices (figure \ref{crys_illust}(c)).

The outline of the present article is as follows: 
We introduce the models in section \ref{models}, calculate the partition function and order parameter in section \ref{calculation}, examine the temperature dependence of the order parameter in section \ref{numerical}, and discuss the prospect of this study in section \ref{discussion}. Summary is given in section \ref{summary}.

\section{Models \label{models}}
In the present article, we consider a binary mixture composed of two types of particles called A and B, and the Hamiltonian is given by
\begin{eqnarray}
 H_{ \left\{ \vector{k}_{\alpha} \right\} } & = & \sum _{i} \frac{\vector{p}_i ^2}{2m_A} +  \sum _{I} \frac{\vector{p}_I ^2}{2m_B} 
 - \frac{J_{AA} }{N} \sum _{\alpha }  \sum _{i,j} \cos \vector{k}_{\alpha } \cdot ( \vector{x}_i - \vector{x}_j ) \nonumber \\ & & - \frac{J_{AB} }{N} \sum _{\alpha }  \sum _{i,J} \cos \vector{k}_{\alpha } \cdot ( \vector{x}_i - \vector{x}_J + \vector{\phi} ) - \frac{J_{BB} }{N} \sum _{\alpha }  \sum _{I,J} \cos \vector{k}_{\alpha } \cdot ( \vector{x}_I - \vector{x}_J ) \nonumber \\
& &- \sum _{\alpha } h_{A \alpha} \sum _i \cos \vector{k}_{\alpha } \cdot \vector{x}_i - \sum _{\alpha } h_{B \alpha} \sum _I \cos \vector{k}_{\alpha } \cdot \left( \vector{x}_I- \vector{\phi} \right) ,
 \label{Hamiltonian1}
\end{eqnarray}
where the lower-case suffix ($i, j$) and upper-case one ($I, J$) represents particles A and B, respectively. Furthermore, $\vector{\phi}$ is a constant vector which gives the distance between particles A and B in the perfect crystal (see figure \ref{hl_illust}), and $\left\{ \vector{k}_{\alpha } \right\}$ is the set of smallest reciprocal lattice vectors of the crystal structure under consideration. A pair of reciprocal lattice vectors  $\pm \vector{k}_{\alpha } $ are regarded as one vector as in our previous study. The number of particles A and B are represented by $N_A$ and $N_B$, and the number of whole particles by $N = N_A + N_B$. The weak positive external field $h_{A \alpha} $ and $h_{B \alpha} $ are introduced in order to treat the symmetry breaking by Bogoliubov quasi-average method\cite{Bogoliubov}. For simplicity, we assume that the coupling constants $J_{AA}$, $J_{BB}$ and $J_{AB}$ have the positive value.

Each particle forms a sublattice of the crystal. For example, in order to construct the hexagonal lattice, we consider the model in which each type of particles form the triangular lattice, namely the sublattice of the hexagonal one. Next, we introduce  the interaction between different types of particles so that the distance between two sublattices coincides with the crystal realized in the solid phase of the original model as shown in figure \ref{hl_illust}. 
Here we adopt the sublattice structures in which one position is allocated for one primitive cell and the reciprocal lattice spaces are spanned by the smallest reciprocal lattice vectors~\cite{KH}. Note that the crystal structures depend on the choice of $\left\{ \vector{k}_{\alpha } \right\}$; when we choose one set of $\left\{ \vector{k}_{\alpha } \right\}$, only one corresponding structure is stabilized.

To express the following calculation simply, we put $\vector{x} ' _I \equiv \vector{x}_I - \vector{\phi}$, 
as new $\vector{x} _I$ and delete $\vector{\phi}$ from ~(\ref{Hamiltonian1}) as 
\begin{eqnarray}
 H_{ \left\{ \vector{k}_{\alpha} \right\} } & = & \sum _{i} \frac{\vector{p}_i ^2}{2m_A} +  \sum _{I} \frac{\vector{p}_I ^2}{2m_B} 
 - \frac{J_{AA} }{N} \sum _{\alpha }  \sum _{i,j} \cos \vector{k}_{\alpha } \cdot ( \vector{x}_i - \vector{x}_j ) \nonumber \\ & & - \frac{J_{AB} }{N} \sum _{\alpha }  \sum _{i,J} \cos \vector{k}_{\alpha } \cdot ( \vector{x}_i - \vector{x}_J) - \frac{J_{BB} }{N} \sum _{\alpha }  \sum _{I,J} \cos \vector{k}_{\alpha } \cdot ( \vector{x}_I - \vector{x}_J ) \nonumber \\
& &- \sum _{\alpha } h_{A \alpha} \sum _i \cos \vector{k}_{\alpha } \cdot \vector{x}_i - \sum _{\alpha } h_{B \alpha} \sum _I \cos \vector{k}_{\alpha } \cdot \vector{x}_I.
 \label{Hamiltonian1.5}
\end{eqnarray}
Note that this transformed Hamiltonian depends only on the form of sublattice, hence the behavior of thermodynamic quantities of NaCl-type and sphalerite-type crystals are the same under this model.

\section{Calculation of the partition function \label{calculation}}
In the present section, we calculate partition function of the model:
\begin{eqnarray}
Z_{ \left\{ \vector{k}_{\alpha} \right\} } & \equiv & \frac{1}{h^{d(N_A +N_B)} N_A ! N_B !} \int  _{\Omega } \prod _i d \vector{x} _i  \int \prod _i d \vector{p} _i \exp \left( - \beta  H_{ \left\{ \vector{k}_{\alpha} \right\} } \right) \nonumber \\
 & = & \frac{1}{\Lambda _A ^{dN_A} \Lambda _B ^{dN_B} N_A ! N_B !} \int  _{\Omega } \prod _i d \vector{x} _i  \cdot \nonumber \\ 
& & \exp \left( \frac{\beta J_{AA}}{N} \sum _{\alpha } \sum _{i,j} \cos \vector{k} _{\alpha } \cdot ( \vector{x} _i - \vector{x} _j ) \right. \nonumber \\ 
& & \hspace{10mm} + \frac{\beta J_{AB}}{N} \sum _{\alpha } \sum _{i,J} \cos \vector{k} _{\alpha } \cdot ( \vector{x} _i - \vector{x} _J ) + \frac{\beta J_{BB}}{N} \sum _{\alpha } \sum _{I,J} \cos \vector{k} _{\alpha } \cdot ( \vector{x} _I - \vector{x} _J ) \nonumber \\
& & \hspace{10mm} \left. + \beta  \sum _{\alpha } h _{A \alpha } \sum _i \cos \vector{k} _{\alpha } \cdot \vector{x} _i + \beta \sum _{\alpha } h _{B \alpha } \sum _I \cos \vector{k} _{\alpha } \cdot \vector{x} _I \right)
\label{Z0}
\end{eqnarray}
\begin{eqnarray}
\mathrm{with} \ \ \frac{1}{\Lambda _A} = \frac{1}{h} \sqrt{\frac{2 \pi m_A}{\beta }} , \ \ \frac{1}{\Lambda _B} = \frac{1}{h} \sqrt{\frac{2 \pi m_B}{\beta }} .
\end{eqnarray}
First, cosine potentials are transformed to one-body potentials with auxiliary variables by Hubbard-Stratonovich transformation:
\begin{eqnarray}
& & \exp \left( \frac{\beta J_{AA}}{N} \sum _{\alpha } \sum _{i,j} \cos \vector{k} _{\alpha } \cdot ( \vector{x} _i - \vector{x} _j ) + \cdots \right) \nonumber \\
& = & \exp \left[ \sum _{\alpha } \left\{ \frac{\beta J_{AA}}{N} \left( \sum _i \cos \vector{k} _{\alpha } \cdot \vector{x} _i \right) ^2 + \frac{\beta J_{AA}}{N} \left( \sum _i \sin \vector{k} _{\alpha } \cdot \vector{x} _i \right) ^2 \right. \right. \nonumber \\
& & \hspace{15mm} + \frac{\beta J_{AB}}{N} \left( \sum _i \cos \vector{k} _{\alpha } \cdot \vector{x} _i \right) \cdot \left( \sum _I \cos \vector{k} _{\alpha } \cdot \vector{x} _I \right) \nonumber \\
& & \hspace{15mm} + \frac{\beta J_{AB}}{N} \left( \sum _i \sin \vector{k} _{\alpha } \cdot \vector{x} _i \right) \cdot \left( \sum _I \sin \vector{k} _{\alpha } \cdot \vector{x} _I \right) \nonumber \\
& & \hspace{15mm} + \frac{\beta J_{BB}}{N} \left( \sum _I \cos \vector{k} _{\alpha } \cdot \vector{x} _I \right) ^2 + \frac{\beta J_{BB}}{N} \left( \sum _I \sin \vector{k} _{\alpha } \cdot \vector{x} _I \right) ^2 \nonumber \\
& & \hspace{15mm} + \left. \left. \beta  \sum _{\alpha } h _{A \alpha } \sum _i \cos \vector{k} _{\alpha } \cdot \vector{x} _i + \beta \sum _{\alpha } h _{B \alpha } \sum _I \cos \vector{k} _{\alpha } \cdot \vector{x} _I  \right\} \right] \nonumber \\
& = & \int \prod _{\alpha } \left\{ \left( \frac{\beta N}{\pi} \right) ^4 J_{AA} J_{AB} ^2 J_{BB} dq_{Ac \alpha } dq_{As \alpha } dq_{Bc \alpha } dq_{Bs \alpha } dz_{Rc \alpha} dz_{Ic \alpha} dz_{Rs \alpha} dz_{Is \alpha} \right\} \nonumber \\
& & \cdot \exp \left[ \sum _{\alpha } \left\{ - \beta J_{AA} N ( q _{Ac \alpha } ^2 +q _{As \alpha } ^2 ) - \beta J_{BB} N ( q _{Bc \alpha } ^2 +q _{Bs \alpha } ^2 ) - \beta J_{AB} N ( | z _{c \alpha } | ^2 +| z _{s \alpha } | ^2 ) \right. \right. \nonumber \\
 & & \hspace{16mm}  + \beta ( 2 J_{AA} q_{Ac \alpha } + J_{AB} z_{c \alpha} + h_{\alpha }) \sum _i \cos \vector{k} _{\alpha } \cdot \vector{x} _i  \nonumber \\
& & \hspace{16mm}  + \beta ( 2 J_{BB} q_{Bc \alpha } + J_{AB} \bar{z} _{c \alpha} + h_{\alpha }) \sum _I \cos \vector{k} _{\alpha } \cdot \vector{x} _I \nonumber \\
 & &  \hspace{16mm}  + \beta ( 2 J_{AA} q_{As \alpha } + J_{AB} z_{s \alpha}) \sum _i \sin \vector{k} _{\alpha } \cdot \vector{x} _i \nonumber \\
 & &  \hspace{16mm} + \left. \left.  \beta ( 2 J_{BB} q_{Bs \alpha } + J_{AB} \bar{z} _{s \alpha} ) \sum _I \sin \vector{k} _{\alpha } \cdot \vector{x} _I \right\} \right],
\end{eqnarray}
\begin{equation}
\mathrm{with} \ \ z_{c \alpha} = z_{Rc \alpha}+ iz_{Ic \alpha}, \ \ z_{s \alpha} = z_{Rs \alpha} + iz_{Is \alpha}.
\end{equation}
Using the saddle-point method to these auxiliary variables, the partition function is written as
\begin{eqnarray}
Z_{ \left\{ \vector{k}_{\alpha} \right\} } 
& \simeq & \frac{1}{\Lambda _A ^{dN_A} \Lambda _B ^{dN_B} N_A ! N_B !} \cdot \sum _{\vector{q}, \vector{z}:\mathrm{saddle \ points} } \cdot  \nonumber \\ 
 & & \left[ \exp \left\{  - \beta N \sum _{\alpha} \left\{ J_{AA} (q _{Ac \alpha} ^2 + q _{As \alpha} ^2 )  + J_{BB} ( q _{Bc \alpha } ^2 +q _{Bs \alpha } ^2 ) + J_{AB} ( | z _{c \alpha } | ^2 +| z _{s \alpha } | ^2 )  \right\} \right\} \right. \nonumber \\ 
& & \cdot \left[ \int _{\Omega } d \vector{x} \cdot \exp \left\{ \sum _{\alpha } \left\{ \beta ( 2 J_{AA} q_{Ac \alpha } + J_{AB} z_{c \alpha} + h_{A \alpha }) \cos \vector{k} _{\alpha } \cdot \vector{x} \ \right. \right. \right. \nonumber \\
 & &  \hspace{30mm} \Biggl. \Biggl. \left. + \beta ( 2 J_{AA} q_{As \alpha } + J_{AB} z_{s \alpha}) \sin \vector{k} _{\alpha } \cdot \vector{x}  \right\} \Biggr\} \Biggr] ^{N_A}  \nonumber \\
& & \left. \cdot
\left[ \int _{\Omega } d \vector{x} \cdot \exp \left\{ \sum _{\alpha } \left\{ \beta ( 2 J_{BB} q_{Bc \alpha } + J_{AB} \bar{z} _{c \alpha} + h_{B \alpha }) \cos \vector{k} _{\alpha } \cdot \vector{x} \right. \right. \right. \right. \nonumber \\
 & &  \hspace{30mm} \Biggl. \Biggl. \Biggl. \left. + \beta ( 2 J_{BB} q_{Bs \alpha } + J_{AB} \bar{z} _{s \alpha} ) \sin \vector{k} _{\alpha } \cdot \vector{x}  \right\} \Biggr\} \Biggr] ^{N_B} \Biggr]  .  
\label{Z1}
\end{eqnarray}
Here the summation is taken over the saddle points of the function inside of the outer square bracket reinterpreted as the complex function of $8 n_{\alpha}$ complex variables, $q _{Ac \alpha}$, $q _{As \alpha}$,$\ldots$, $z_{Ic \alpha}$ and $z_{Is \alpha}$. This equation becomes exact in the limit of $N \rightarrow \infty$, in which contribution of the points other than the saddle points are ignorable.

We define $ V_{\mathrm{p}}$ and $ \Phi _{\mathrm{p}}$ as the integrals over a primitive cell~\cite{KH}:
\begin{eqnarray}
 \Phi _{\mathrm{p} } \left( \left\{ r_{\alpha} \right\}, \left\{ r' _{\alpha} \right\}, \left\{ \vector{k}_{\alpha} \right\}  \right) & \equiv & \int _{\mathrm{primitive \ cell} } d \vector{x}  \cdot \exp \left\{ \sum _{\alpha } \left( \beta r_{\alpha } \cos \vector{k} _{\alpha } \cdot \vector{x} + \beta r '_{\alpha } \sin \vector{k} _{\alpha } \cdot \vector{x} \right) \right\} \label{Kpr} , \\
 V_{\mathrm{p} }& \equiv & \int _{\mathrm{primitive \ cell} } d \vector{x}  \cdot 1 \label{Vpr} ,
\end{eqnarray}
and $\exp f \left( \vector{q} , \vector{z} \right) $ as
\begin{eqnarray}
& & \exp f \left(  \vector{q} , \vector{z}  \right)  \nonumber \\
 & \equiv & \exp \left\{ - \beta N \sum _{\alpha} \left( J_{AA} (q _{Ac \alpha} ^2 + q _{As \alpha} ^2 )  + J_{BB} ( q _{Bc \alpha } ^2 +q _{Bs \alpha } ^2 ) + J_{AB} ( | z _{c \alpha } | ^2 +| z _{s \alpha } | ^2 ) \right) \right\}  \nonumber \\ 
 & &  \left. \left. \cdot  \Phi _{\mathrm{p} } \left( \left\{  2 J_{AA} q_{Ac \alpha } + J_{AB} z_{c \alpha} + h_{A \alpha } \right\}, \left\{  2 J_{AA} q_{As \alpha } + J_{AB} z_{s \alpha} \right\}, \left\{ \vector{k}_{\alpha} \right\} \right) ^{N_A} \right. \right. \nonumber \\
 & & \hspace{1.5mm} \Biggl. \cdot \Phi _{\mathrm{p} } \left( \left\{ 2 J_{BB} q_{Bc \alpha } + J_{AB} \bar{z}_{c \alpha} + h_{B \alpha } \right\}, \left\{  2 J_{BB} q_{Bs \alpha } + J_{AB} \bar{z}_{s \alpha} \right\}, \left\{ \vector{k}_{\alpha} \right\} \right) ^{N_B} \label{fpr} .
\end{eqnarray}
Using the periodicity of the integrands, the partition function is given by
\begin{eqnarray}
Z_{ \left\{ \vector{k}_{\alpha} \right\} } & \simeq & \frac{1}{\Lambda _A ^{dN_A} \Lambda _B ^{dN_B} N_A ! N_B !} \left( \frac{V}{V_p} \right) ^N  \sum _{\vector{q}, \vector{z}:\mathrm{saddle \ point} } \exp f \left(  \vector{q} , \vector{z}  \right)   \nonumber \\
 & \simeq & \frac{1}{\Lambda _A ^{dN_A} \Lambda _B ^{dN_B} N_A ! N_B !} \left( \frac{V}{V_p} \right) ^N \cdot \exp f \left( \vector{q_0}, \vector{z_0} \right)  . \nonumber \\
\label{Z1.5}
\end{eqnarray}
Here the set of complex variables $( \vector{q_0}, \vector{z_0} )$ means the saddle point which gives the largest $\left| \exp f \right|$ among all saddle points. Contribution of other saddle points is ignorable when $N$ is large. 


The saddle-point condition is expressed as the following four self-consistent equations:
\begin{eqnarray}
& & \frac{\partial f}{\partial q_{Ac \alpha}} = 0 \nonumber \\
 & \Leftrightarrow & \ \ 2 \beta N q_{Ac \alpha} \nonumber \\
&  & = 2 N_A \left. \frac{\partial}{\partial r_{\alpha}} \log \Phi _{\mathrm{p} } \left( \left\{ r_{\alpha} \right\},  \left\{ r'_{\alpha} \right\} , \left\{ \vector{k}_{\alpha} \right\}  \right) \right| _{r_{\alpha} =  2 J_{AA} q_{Ac \alpha } + J_{AB} z_{c \alpha} + h_{A \alpha }, r' _{\alpha} =  2 J_{AA} q_{As \alpha } + J_{AB} z_{s \alpha}} \label{SC1} \\
& & \frac{\partial f}{\partial q_{Bc \alpha}} = 0 \nonumber \\
 & \Leftrightarrow & \ \ 2 \beta N q_{Bc \alpha} \nonumber \\
 &  & = 2 N_B \left. \frac{\partial}{\partial r_{\alpha}} \log \Phi _{\mathrm{p} } \left( \left\{ r_{\alpha} \right\} , \left\{ r'_{\alpha} \right\}, \left\{ \vector{k}_{\alpha} \right\}  \right) \right| _{r_{\alpha} =  2 J_{BB} q_{Bc \alpha } + J_{AB} \bar{z}_{c \alpha} + h_{B \alpha },r'_{\alpha} =  2 J_{BB} q_{Bs \alpha } + J_{AB} \bar{z}_{s \alpha}} \label{SC2} 
\end{eqnarray}
\begin{eqnarray}
& & \frac{\partial f}{\partial z_{Rc \alpha}} = 0 \nonumber \\
& \Leftrightarrow & \ \ 2 \beta N z_{Rc \alpha} \nonumber \\
 & \ & =  N_A \left. \frac{\partial}{\partial r_{\alpha}} \log \Phi _{\mathrm{p} } \left( \left\{ r_{\alpha} \right\}, \left\{ r'_{\alpha} \right\} , \left\{ \vector{k}_{\alpha} \right\}  \right) \right| _{r_{\alpha} =  2 J_{AA} q_{Ac \alpha } + J_{AB} z_{c \alpha} + h_{A \alpha }, r' _{\alpha} =  2 J_{AA} q_{As \alpha } + J_{AB} z_{s \alpha}} \nonumber \\
 & \ & \hspace{3mm} + N_B \left. \frac{\partial}{\partial r_{\alpha}} \log \Phi _{\mathrm{p} } \left( \left\{ r_{\alpha} \right\}, \left\{ r'_{\alpha} \right\} , \left\{ \vector{k}_{\alpha} \right\}  \right) \right| _{r_{\alpha} =  2 J_{BB} q_{Bc \alpha } + J_{AB} \bar{z}_{c \alpha} + h_{B \alpha },r'_{\alpha} =  2 J_{BB} q_{Bs \alpha } + J_{AB} \bar{z}_{s \alpha}} \label{SC3} \\
& & \frac{\partial f}{\partial z_{Ic \alpha}} = 0 \nonumber \\
& \Leftrightarrow & \ \ 2 \beta N z_{Ic \alpha} \nonumber \\ 
& \ & =  iN_A \left. \frac{\partial}{\partial r_{\alpha}} \log \Phi _{\mathrm{p} } \left( \left\{ r_{\alpha} \right\}, \left\{ r'_{\alpha} \right\} , \left\{ \vector{k}_{\alpha} \right\}  \right) \right| _{r_{\alpha} =  2 J_{AA} q_{Ac \alpha } + J_{AB} z_{c \alpha} + h_{A \alpha }, r' _{\alpha} =  2 J_{AA} q_{As \alpha } + J_{AB} z_{s \alpha}} \nonumber \\
 & \ & \hspace{3mm} - iN_B \left. \frac{\partial}{\partial r_{\alpha}} \log \Phi _{\mathrm{p} } \left( \left\{ r_{\alpha} \right\}, \left\{ r'_{\alpha} \right\} , \left\{ \vector{k}_{\alpha} \right\}  \right) \right| _{r_{\alpha} =  2 J_{BB} q_{Bc \alpha } + J_{AB} \bar{z}_{c \alpha} + h_{B \alpha },r'_{\alpha} =  2 J_{BB} q_{Bs \alpha } + J_{AB} \bar{z}_{s \alpha}} \label{SC4} 
\end{eqnarray}
From these equations, we obtain
\begin{equation}
z_{Rc \alpha} = \frac{q_{Ac \alpha } + q_{Bc \alpha } }{2} , \ \mathrm{and} \ \ z_{Ic \alpha} = i \cdot \frac{q_{Ac \alpha } - q_{Bc \alpha } }{2} , \label{SC3.5}
\end{equation}
 i.e. $z_{c \alpha} = q_{Bc \alpha }$, and $\bar{z}_{c \alpha} = q_{Ac \alpha }$. Similar relations also exist in terms of $z_{s \alpha}$ and $\bar{z}_{s \alpha}$. Namely, the following relations exist between the variables appeared in (\ref{Z1.5}):
\begin{eqnarray}
z_{c \alpha} =q_{0Bc \alpha }, z_{s \alpha} =q_{0Bs \alpha }, \bar{z} _{c \alpha } = q_{0Ac \alpha }, \ \mathrm{and} \ \bar{z} _{s \alpha } = q_{0As \alpha }  .
\label{Z2} 
\end{eqnarray}
Here, $q_{0A \alpha}$ and $q_{0B \alpha}$ are given as solutions of self-consistent equations (\ref{SC1}) and (\ref{SC2}).

The order parameters of this system are given by the following two equations:
\begin{eqnarray}
 \left< \rho _{\vector{k} _{A \alpha} } \right> & \equiv & \left< \frac{1}{N} \sum _{i} \cos \vector{k} _{\alpha} \cdot \vector{x} _i  \right> = \left. \frac{1}{\beta N} \frac{\partial}{\partial h_{A \alpha}} \log Z \right| _{ h_{\alpha} \rightarrow +0} \nonumber \\
 \left< \rho _{\vector{k} _{B \alpha} } \right> & \equiv & \left< \frac{1}{N} \sum _{I} \cos \vector{k} _{\alpha} \cdot \vector{x} _I  \right> = \left. \frac{1}{\beta N} \frac{\partial}{\partial h_{B \alpha}} \log Z \right| _{ h_{\alpha} \rightarrow +0} . \label{op1} 
\end{eqnarray}
Calculating the right-hand side of (\ref{op1}) using the self-consistent equations, we find that these parameters coincide with the auxiliary variables
\begin{eqnarray}
\left< \rho _{\vector{k} _{A \alpha} } \right> & = & \lim _{ h_{A \alpha} , h_{B \alpha} \rightarrow +0 }  \frac{N_A}{\beta N} \cdot \nonumber \\ & & \left. \frac{\partial}{\partial r_{\alpha}} \log \Phi _{\mathrm{p} } \left( \left\{ r_{\alpha} \right\} , \left\{ r'_{\alpha} \right\} , \left\{ \vector{k}_{\alpha} \right\}  \right) \right| _{r_{\alpha} =  2 J_{AA} q_{0Ac \alpha } + J_{AB} q_{0Bc \alpha} + h_{A \alpha }, r'_{\alpha} =  2 J_{AA} q_{0As \alpha } + J_{AB} q_{0Bs \alpha} } \nonumber \\
& = & \lim _{ h_{A \alpha} , h_{B \alpha} \rightarrow +0}  q_{0Ac \alpha } \ , \nonumber \\
 \left< \rho _{\vector{k} _{B \alpha} } \right> & = & \lim _{ h_{A \alpha} , h_{B \alpha} \rightarrow +0}  q_{0Bc \alpha } \ . \label{op2} 
\end{eqnarray}
Similarly, we obtain
\begin{eqnarray}
 \left< \frac{1}{N} \sum _{i} \sin \vector{k} _{\alpha} \cdot \vector{x} _i  \right>  =   q_{0As \alpha } , \ \mathrm{and} \ \ \left< \frac{1}{N} \sum _{I} \sin \vector{k} _{\alpha} \cdot \vector{x} _I  \right> = q_{0Bs \alpha }  . \label{op_sin} 
\end{eqnarray}
However, these variables are zero because of the existence of external field $h_{A \alpha}$ and $ h_{B \alpha}$ \cite{KH}.
Note that we define the solid phase as the state in which $\left< \right. \rho _{\vector{k} _{A \alpha} } \left. \right> $ or $\left< \right. \rho _{\vector{k} _{B \alpha} } \left. \right> $ takes a nonzero value. Such definition, first introduced by Kirkwood and Monroe~\cite{KirMon} and Vlasov~\cite{Bazarov} is useful when we regard the solid-fluid phase transition as the breaking of continuous translational symmetry, although there is an alternative definition of the solid phase by the existence of the shear modulus~\cite{MBorn}.

At the end of this section, we prove that the transition points for two order parameters $\left< \right. \rho _{\vector{k} _{A \alpha} } \left. \right> $ and $\left< \right. \rho _{\vector{k} _{B \alpha} } \left. \right> $, or the temperatures at which they change from zero to nonzero, coincide unless $J_{AB}=0$ in the limit of $ h_{A \alpha}, h_{B\alpha} \rightarrow +0$. We assume the existence of the state where one of these two parameter is nonzero, and the other is zero, and derive the contradiction. In this discussion, we can let particle A has the nonzero order parameter, i.e. $q _{A \alpha} > 0 $, and $ q _{B \alpha} = 0$, without losing generality. Equation (\ref{SC2}) is expressed as
\begin{equation}
 0 = \left. \frac{\partial}{\partial r_{\alpha}} \log \Phi _{\mathrm{p} } \left( \left\{ r_{\alpha} \right\} , \left\{ 0 \right\} , \left\{ \vector{k}_{\alpha} \right\}  \right) \right| _{r_{\alpha} =  J_{AB} q_{A \alpha} } \label{SC2_B0} 
\end{equation}
under this assumption. According to the power series expansion of $\Phi _{\mathrm{p}}$, however, the coefficient for each power of $r _{\alpha}$ has positive value. (See section 4 of \cite{KH}.) Hence, the right-hand side of (\ref{SC2_B0}) should be positive, and it contradicts the assumption. We conclude from this fact that $q _{A \alpha} $ and $ q _{B \alpha}$ simultaneously change from zero to nonzero.

\section{Behavior of the order parameter \label{numerical}}
\begin{figure}[!hbp]
\begin{center}
\includegraphics[width = 15.0cm]{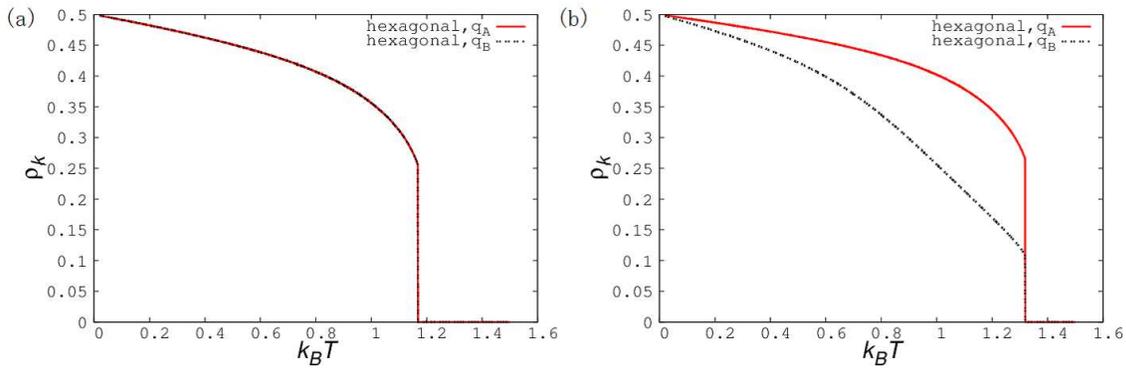}
\caption{Temperature dependence of $q_{A }$ and $q_{B }$ for the hexagonal lattice. (a) $J_{AA}=J_{BB}=J_{AB}=4/3$. (b) $J_{AA}=2, J_{BB}=2/3, J_{AB}=4/3$. }
\label{F_HL}
\end{center}
\end{figure}

\begin{figure}[!hbp]
\begin{center}
\includegraphics[width = 15.0cm]{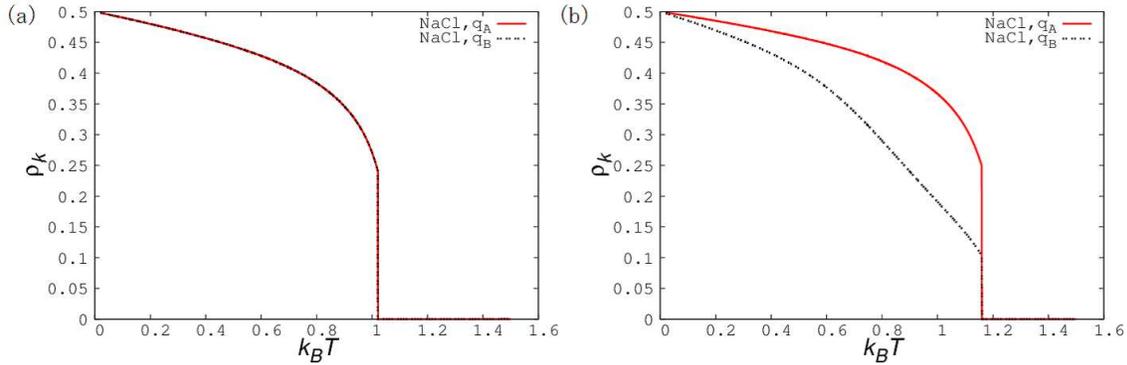}
\caption{Temperature dependence of $q_{A }$ and $q_{B }$ for the NaCl-type (or sphalerite-type) crystal. (a) $J_{AA}=J_{BB}=J_{AB}=4/3$. (b)  $J_{AA}=2, J_{BB}=2/3, J_{AB}=4/3$. }
\label{F_NaCl}
\end{center}
\end{figure}

\begin{figure}[!hbp]
\begin{center}
\includegraphics[width = 15.0cm]{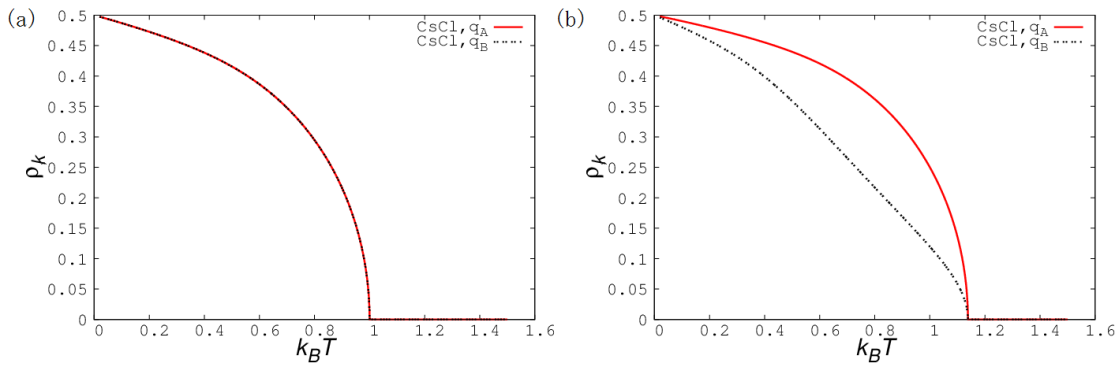}
\caption{Temperature dependence of $q_{A }$ and $q_{B }$ for the CsCl-type crystal. (a) $J_{AA}=J_{BB}=J_{AB}=4/3$. (b) $J_{AA}=2, J_{BB}=2/3, J_{AB}=4/3$. }
\label{F_CsCl}
\end{center}
\end{figure}

In this section, the temperature dependence of the order parameters  $\left< \right. \rho _{\vector{k} _{A \alpha} } \left. \right> $ and $\left< \right. \rho _{\vector{k} _{B \alpha} } \left. \right> $ is calculated by evaluating (\ref{Z1.5}) numerically. This partition function is independent of $k$, or the length of $\vector{k}_{\alpha}$~\cite{KH}. Hence we let $k=1$, and also assume that the values of  $\left< \right. \rho _{\vector{k} _{A \alpha} } \left. \right> $ and $\left< \right. \rho _{\vector{k} _{B \alpha} } \left. \right> $ are given independently of $\alpha$ as $q_A$ and $q_B$. We investigate hexagonal lattice, NaCl-type (or sphalerite-type), and CsCl-type crystal for the two set of parameters with $J_{AA}=J_{BB}=J_{AB}=4/3$, and $J_{AA}=2, J_{BB}=2/3, J_{AB}=4/3$. The number of the particles for each sublattice is set to be equal: $N_A = N_B$. The results are shown in figures \ref{F_HL}--\ref{F_CsCl}.

 In these figures, $q _{A \alpha} $ and $ q _{B \alpha}$ changes from zero to nonzero at the same temperature, which is consistent with our argument in the previous section. If $J_{AA}=J_{BB} , N_A = N_B ,$ the values of $q_A$ and $q_B$ coincide with each other because of the symmetry between two sublattices. The phase transition is of first order when the set of $\left\{ \vector{k}_{\alpha } \right\} $ are linearly dependent, and of second order when they are linearly dependent. In the case of the crystal structures we treat in this section, the model of CsCl-type crystal shows the second-order phase transition, while that of hexagonal lattice and NaCl-type (or sphalerite-type) crystal shows the first-order one. Such relation between the order of phase transitions and the set of $\left\{ \vector{k}_{\alpha } \right\} $ is the same as that in simpler crystals~\cite{KH}. Note that the order of the phase transition, or whether the wave numbers of the cosine potentials are linearly independent or not, is determined only by the algebraic relation between these wave numbers, and does not depend on sizes or angles of them.

\section{Discussion \label{discussion}}
In the present paper, we have studied the model in which crystal structures are composed of two sublattices. Although we have restricted the number of sublattices as two in this paper, expanding our method to more complicated structures containing three or more sublattices may be straightforward. However, the sublattices treated in the present paper are limited to those of whose reciprocal space can be spanned by their smallest reciprocal lattice vectors~\cite{KH, Kittel}, whereas sublattices of several types of real crystal structures such as hcp cannot. It is also difficult to treat the case where sublattices A and B are composed of the same kind of particles. For example, the positions of the particles in the diamond-type crystal are the same as those of sphalerite-type crystal, but we can not use the method of this paper to it.

\begin{figure}[!hbp]
\begin{center}
\includegraphics[width = 8.0cm]{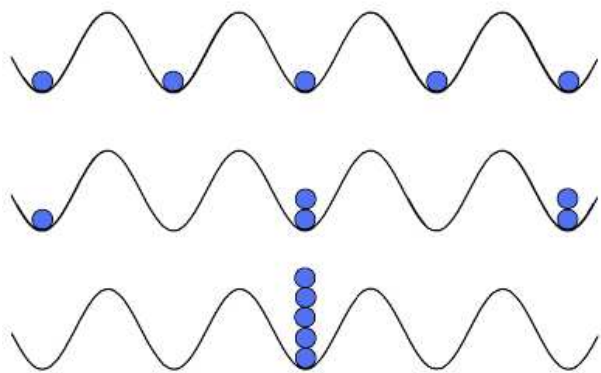}
\caption{The feature of cosine potential: Although cosine potentials gather particles on particular lattice points, they do not care which one among these points the particles come. For example, the system has the same energy in the above three cases. }
\label{ig_illust}
\end{center}
\end{figure}

Similarly to the simple crystals~\cite{KH}, the present system obeys the ideal-gas law even in the solid phase: 
\begin{equation}
 P  = \frac{\partial}{\partial V} \left( \frac{\log Z}{\beta} \right) = \frac{N}{\beta V} .
\end{equation}
The origin of this unphysical behavior is the periodicity of the cosine potentials. They have the same value no matter which lattice points the particles occupy as shown in figure \ref{ig_illust}, therefore, the number of particles existing at each lattice points has no restriction.

Although these problems exist, the trial to find relationship between this model and real solid-fluid phase transition seems to be valuable because there are few theoretical description which treat both of the solid and fluid phase by one partition function, starting from exactly-solvable models. 

The model treated in the present paper resembles to infinite-range interaction models, or mean-field models of spin systems, in the point that the interaction does not decay with distance and the partition function can be calculated exactly by similar mathematical methods. Hence, this model should be called as a ``mean-field model'' of solid-fluid transition.
 In the case of simple crystal structures, we are now formulating the ``mean-field approximation'' for more realistic models such as the Lennard-Jones system, in order to investigate the relation between the transition of this ``mean-field model'' and real systems~\cite{KH2}. 

\section{Summary \label{summary}}

In this paper, we introduced an exactly-solvable model of solid-fluid phase transitions for crystals which contain two positions for particles in a primitive cell. This is a generalization of our previous scheme which treated simpler ones including only one position for particles in a primitive cell.  Specifically, we consider the binary mixture composed of two types of particles, in which each particle makes a sublattice. The interparticle interaction is given by the sum of cosine potentials whose wave numbers correspond to the smallest reciprocal lattice vectors of these sublattices, and the crystal structure depends on these wave numbers. We studied hexagonal lattice, NaCl-type or sphalerite-type crystal, and CsCl-type crystal. Several features are similar to those in our previous paper. The order of the phase transition is determined by whether the set of wave numbers are linearly independent or not, and the system obeys ideal gas law even in the solid phase. This model also has an interesting feature, i.e. the phase transition of each sublattice occurs at the same temperature. Generalization of the scheme proposed in the present paper to more complicated structures, which have three or more positions for particles in a primitive cell, is considered to be straightforward. 

\section*{Acknowledgements}
The author thanks to Koji Hukushima, Masamichi Nishino, and Yoshihiko Nonomura for helpful discussions.


\clearpage

\section*{Reference}
\begin{thebibliography}{99}
\bibitem{LJD}  J.E.Lennard-Jones and A.F.Devonshire, 1939 \textit{Proc.Roy.Soc.A} \textbf{169.938} \ 317 
\bibitem{MOI} H.Mori, H.Okammoto and S.Isa, 1972 \textit{Prog.The.Phys} \textbf{47} \ 4 
\bibitem{KirMon} J.G.Kirkwood and E.Monroe, 1941 \textit{J.Chem.Phys} \textbf{9} \ 514 
\bibitem{RY} T.V.Ramakrishnan and M.Yussouff, 1979 \textit{Phys.Rev.B} \textbf{19} \ 2775 
\bibitem{CA}  W.A.Curtin and N.W.Ashcroft, 1986 \textit{Phys.Rev.Lett} \textbf{56} \ 2775  
\bibitem{DA} A.R.Denton and N.W.Ashcroft, 1989 \textit{Phys.Rev.A} \textbf{39} \ 4701 
\bibitem{CL} P.M.Chaikin and T.C.Lubensky, 1995 \textit{Principles of condensed matter physics} (Cambridge: Cambridge University Press)
\bibitem{FP} S.Fesjian and J.K.Percus, 1990 \textit{J.Stat.Phys} \textbf{60}:659 
\bibitem{CF} H-O.Carmesin and Y.Fan, 1990 \textit{J.Phys.A:Math.Gen}  \textbf{23}:3613 
\bibitem{KH} H.Komatsu, 2015 \textit{J.Stat.Mech}, P08020 
\bibitem{Yussouff} M.Yussouff \textit{Phys.Rev.B}, 1981 \textbf{23} \ 5871 
\bibitem{Bogoliubov} N.N.Bogoliubov 1970 \textit{Lectures on quantum statistics} Vol.2 (Gordon and Breach: New York)
\bibitem{Bazarov} I.P.Bazarov, 1967 \textit{Sov.Phys.J} \textbf{10.2} \ 53
\bibitem{MBorn} M.Born, 1939 \textit{J.Chem.Phys} \textbf{7} \ 591 
\bibitem{Kittel} Kittel, 2005 \textit{Introduction to solid state physics 8th edition} (New York: Wiley)
\bibitem{KH2} H.Komatsu \ \ in preparation
\end{thebibliography}
\end{document}